# Rise of the machines: first year operations of the Robo-AO visible-light laser-adaptive-optics instrument

**Christoph Baranec,**
*Institute for Astronomy, The University of Hawai'i at Manoa, Hilo, HI 96720*
**Reed Riddle,**
*Caltech Optical Observatories, California Institute of Technology, Pasadena, CA 91125*
**Nicholas M. Law,**
*University of North Carolina at Chapel Hill, Chapel Hill, NC 27599*
**A. N. Ramaprakash,**
*Inter-University Centre for Astronomy & Astrophysics, Ganeshkhind, Pune, 411007, India*
**Shriharsh Tendulkar, Kristina Hogstrom, Khanh Bui,**
*Caltech Optical Observatories, California Institute of Technology, Pasadena, CA 91125*
**Mahesh Burse, Pravin Chordia, Hillol Das,**
*Inter-University Centre for Astronomy & Astrophysics, Ganeshkhind, Pune, 411007, India*
**Richard Dekany, Shrinivas Kulkarni,**
*Caltech Optical Observatories, California Institute of Technology, Pasadena, CA 91125*
**Sujit Punnadi,**
*Inter-University Centre for Astronomy & Astrophysics, Ganeshkhind, Pune, 411007, India*
**& Roger Smith**
*Caltech Optical Observatories, California Institute of Technology, Pasadena, CA 91125*

## ABSTRACT

Robo-AO is the first autonomous laser adaptive optics system and science instrument operating on sky. With minimal human oversight, the system robotically executes large scale surveys, monitors long-term astrophysical dynamics and characterizes newly discovered transients, all at the visible diffraction limit. The average target-to-target operational overhead, including slew time, is a mere 86 s, enabling up to ~20 observations per hour. The first of many envisioned systems went live in June 2012, and has since finished 78 nights of science observing at the Palomar Observatory 60-inch (1.5 m) telescope, with over 9,000 robotic observations executed as of August 2013. The system will be augmented in 2014 with a low-noise wide field infrared camera, which will double as a tip-tilt sensor, to widen the spectral bandwidth of observations, increase available sky coverage as well as enable deeper visible imaging using adaptive-optics sharpened infrared tip-tilt guide sources.

## 1. INTRODUCTION

Robo-AO is a new autonomous laser-guide-star adaptive-optics (AO) and science instrument currently deployed on the 60-inch telescope at Palomar Observatory. Preliminary results [1-4] from the prototype system demonstrate visible-light imaging with angular resolutions approaching the diffraction limit of a 1.5-m telescope, ≈ 0.12". Robotic software automations [5] keep target-to-target observing overheads of < 1.5 minutes (including slew time) leading to the observation of over two-hundred targets per night, and the completion of the three largest AO surveys to date. Many other implementations of Robo-AO are in various stages of development: a clone of Robo-AO for the 2-m IGO [6] in India is underway; a natural guide star only system for the 1-m Table Mountain telescope, led by Pomona College, has recently closed their AO loop on sky [7]; and all-sky Robo-AO network is being studied for high-resolution follow-up of current and future exoplanet and transient-search surveys, e.g., the Panoramic Survey Telescope & Rapid Response System (Pan-STARRS), the Transiting Exoplanet Survey Satellite (TESS) and the Large Synoptic Survey Telescope (LSST).

## 2. THE ROBO-AO SYSTEM HARDWARE

The first Robo-AO system has been deployed on the Palomar 60-inch (1.5 m) telescope (Fig. 1). The telescope was upgraded with full robotic control in 2004, permitting local or remote operation via TCP/IP commands [8]. In addition, the telescope uses data from a locally hosted weather station to execute automated routines to ensure telescope safety during adverse weather conditions. For a thorough description of the step-by-step operation of Robo-AO, please see [9].

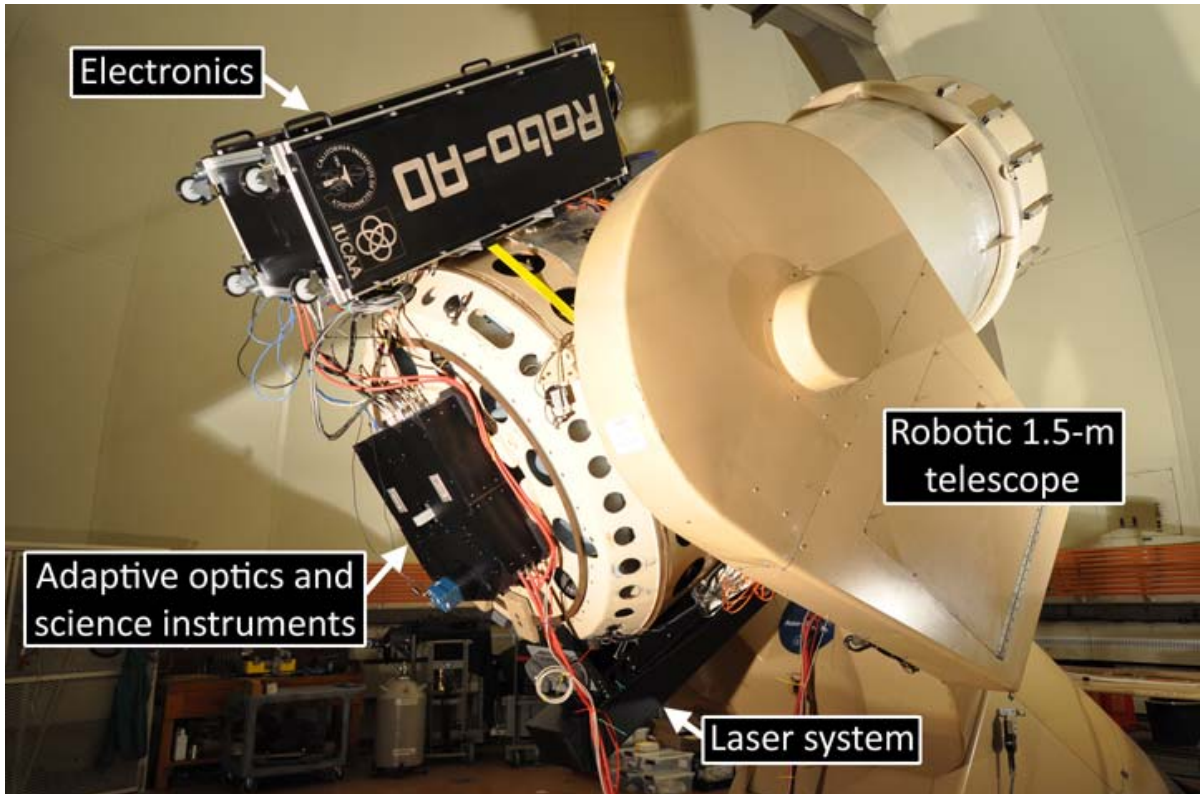

Fig. 1. The Robo-AO laser adaptive-optics system. The adaptive optics and science instruments are installed at the Cassegrain focus of the robotic 60-inch telescope at Palomar Observatory (see Fig. 2). The laser system and support electronics are attached to opposite sides of the telescope tube for balance.

The core of the Robo-AO laser system is a pulsed 10-W ultraviolet laser (JDSU Q301-HD) mounted in an enclosed projector assembly on the side of the telescope. The laser projector incorporates a redundant shutter in addition to the laser's internal shutter for additional safety; a half-wave plate to adjust the angle of projected linear polarization; and an uplink tip-tilt mirror to both stabilize the apparent laser beam position on sky and to correct for telescope flexure. A bi-convex lens on an adjustable focus stage expands the laser beam to fill a 15 cm output aperture lens, which is optically conjugate to the tip-tilt mirror. The output lens focuses the laser light to a line-of-sight distance of 10 km. Within the adaptive-optics instrument, a high-speed $\beta$-$BaB_2O_4$ Pockels cell optical shutter (FastPulse Lasermetrics) is used to transmit laser light only returning from a ~400-m slice of the atmosphere around the 10 km projector focus, resulting in the laser appearing as a spot. Switching of the Pockels cell is driven by the same master clock as the pulsed laser, with a delay to account for the round trip time of the laser pulse through the atmosphere.

The ultraviolet laser has the additional benefit of being invisible to the human eye, primarily due to absorption in the cornea and lens. As such, it is unable to flash-blind pilots and is considered a Class 1 laser system (i.e. incapable of producing damaging radiation levels during operation and exempt from any control measures) for all possible exposures of persons in overflying aircraft, eliminating the need for human spotters located on site as normally required by the Federal Aviation Authority within the U.S.. Unfortunately, the possibility for the laser to damage some satellites in low Earth orbit may exist. For this reason, it is recommended for both safety and liability concerns to coordinate laser activities with an appropriate agency (e.g., with U.S. Strategic Command within the U.S.).

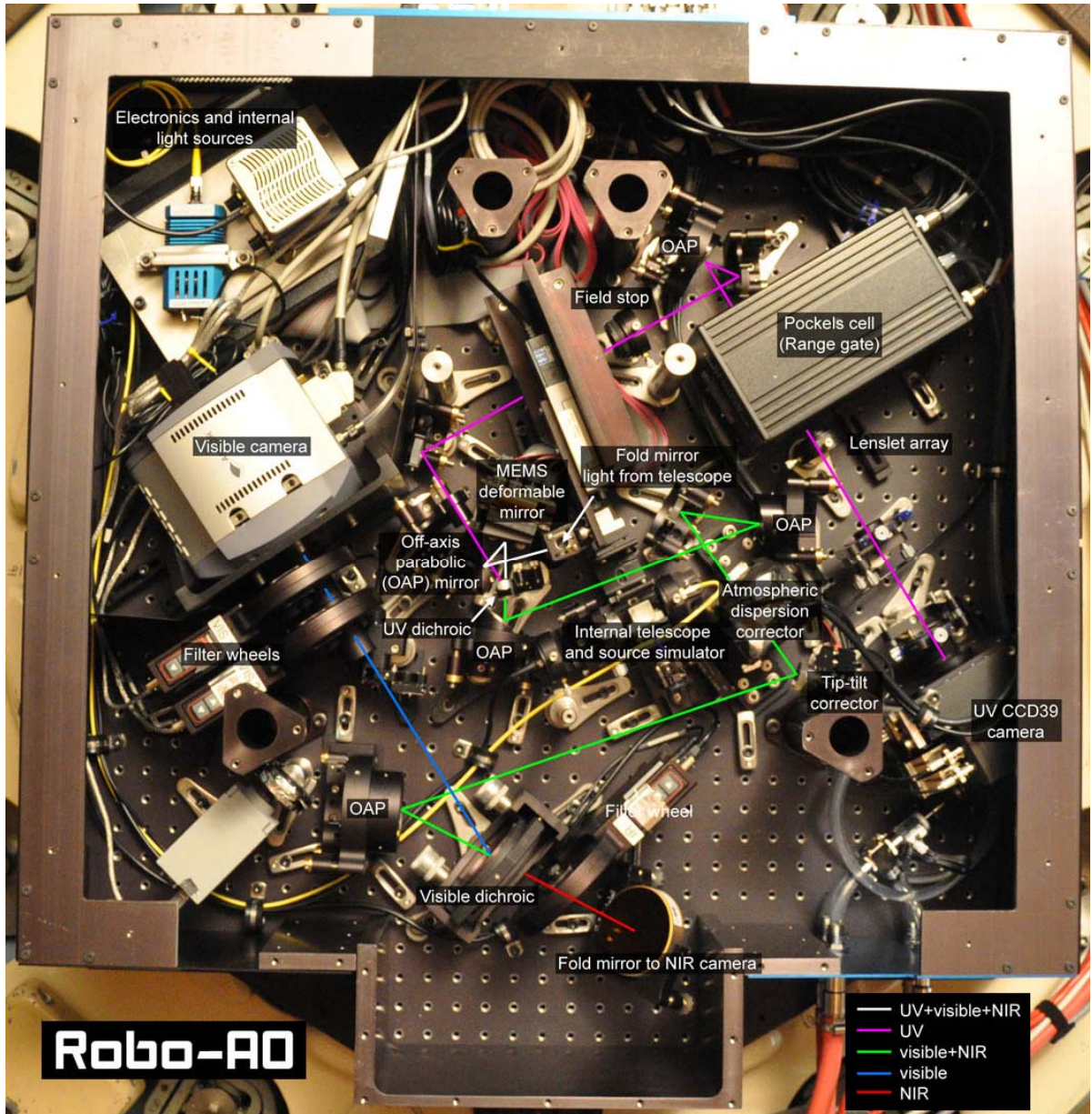

Fig. 2. The Robo-AO adaptive optics system and science instruments. Light focused from the telescope secondary mirror enters through a small hole at the center of the instrument before being reflected by 90 degrees by the first fold mirror towards an off-axis parabolic (OAP) mirror. This mirror images the telescope pupil on the deformable mirror surface. After reflection from the deformable mirror, an UV dichroic splits off the laser light (violet) and directs it to the laser wavefront sensor. An additional reversed OAP mirror within the wavefront sensor corrects the non-common path optical errors introduced by the 10 km conjugate focus of the laser reflecting off of the first OAP mirror. The visible and near-infrared light (green) passing through the UV dichroic is relayed by a pair of OAP mirrors to the atmospheric dispersion corrector. The light is then reflected by the tip-tilt correcting mirror to a final OAP mirror which focuses the light towards the visible dichroic. The visible dichroic reflects the visible light (blue) to the electron-multiplying CCD and transmits the near-infrared light (red) to a fold mirror and ultimately to the infrared camera. The combined UV, visible and near-infrared light from the telescope and source simulator can be directed to the adaptive optics and science instruments by translating the first fold mirror out of the way.

High-order wavefront sensing is performed with an 11×11 Shack-Hartmann wavefront sensor. The detector is an 80×80 pixel format E2V-CCD39 optimized for high quantum efficiency at the laser wavelength (71.9%) and is paired with a set of SciMeasure readout electronics. The detector is binned by a factor of 3. Each Shack-Hartmann subaperture is imaged with 2×2 binned pixels projected to a 5"×5" field of view (limited by a circular 4.8" diameter field stop). The frame transfer time is 500 μs, with frame exposures of 833 μs, corresponding to an effective frame rate of 1.2 kHz. The measured read-noise in this mode is ~6.5 electrons per binned pixel. The laser return signal ranges from 100 to 200 photoelectrons per subaperture per frame (depending on seeing conditions and elevation of observations), equivalent to a signal-to-noise ratio on each slope measurement of 6 to 10.

The wavefront reconstructor matrix is a synthetic modal reconstructor based on the disk harmonic functions up to the 11th radial order, for a total of 75 controlled modes. The AO control loop is based on a simple integral controller with a leaky integrator to a median flat position. Individual modal gain optimization has been performed using on-sky telemetry; however it was found that an overall loop gain of 0.6 was close to optimal in a majority of situations. The closed-loop bandwidth of the system is approximately 90-100 Hz. The tip-tilt modes measured in the wavefront sensor are dominated by mechanical vibration and pointing errors. The tip-tilt signal is used to drive the laser system's uplink tip-tilt mirror, thus keeping the Shack-Hartmann pattern centered on the wavefront sensor.

The high-order wavefront corrector within Robo-AO is a micro-electro-mechanical-systems (MEMS) deformable mirror (Boston Micromachines Multi-DM). Robo-AO uses 120 of the 140 actuators to adjust the illuminated surface of the mirror, sufficient in spatial resolution to accurately fit the calculated correcting shape. The actuators have a maximum surface deviation amplitude of 3.5 μm which corresponds to optical phase compensation of up to 7 μm. In typical seeing at Palomar observatory (median ~1.1"), this compensation length is greater than 5-sigma of the amplitude of the turbulence induced optical error and therefore results in significant correction headroom. Furthermore, the deformable mirror is used to compensate for static optical errors arising from the instrument and telescope at the cost of reduced dynamic range.

Robo-AO uses four off-axis parabolic (OAP) mirrors to relay a 2' diameter full field of view from the telescope to the science cameras achromatically. The relay path includes a fast tip-tilt correcting mirror as well as an atmospheric dispersion corrector (ADC) comprised of a pair of rotating prisms. At the end of the OAP relay is a visible dichroic that reflects light of λ < 950 nm to an electron-multiplying charge-coupled device camera (EMCCD; Andor iXon 888) while transmitting infrared light towards an engineering grade infrared camera (Xenics Xeva-1.7-320). The EMCCD camera has the ability to capture images with very low electronic (detector) noise, at a frame rate which reduces the intra-exposure image motion to below the diffraction-limited angular resolution. By re-centering and stacking a series of these images, a long-exposure image can be synthesized with minimal noise penalty. The EMCCD camera can also be used to stabilize image motion on the infrared camera; measurements of the position of an imaged astronomical source can be used to continuously command the fast tip-tilt to re-point the image to a desired location. Ahead of each camera is a set of filter wheels with an appropriate set of astronomical filters.

An internal telescope and source simulator is integrated into the Robo-AO system as a calibration tool. It can simultaneously simulate the ultraviolet laser focus at 10 km and a blackbody source at infinity, matching the host telescope's focal ratio and exit pupil position. The first fold mirror within Robo-AO directs all light from the telescope's secondary mirror to the adaptive-optics system. The fold mirror is also mounted on a motorized stage which can be translated out of the way to reveal the internal telescope and source simulator.

## 3. VISIBLE-LIGHT ADAPTIVE OPTICS

Individual Robo-AO observations are made with an electron multiplying CCD camera with a 44" square field of view and 0.043" pixel scale. The camera is read out continually at a frame rate of 8.6 Hz during science observations, allowing image motion (which cannot be measured using the laser system) to be removed in software after observations with the presence of a mV ≤ 16 guide star within the field of view. A data reduction pipeline [10] corrects each of the recorded frames for detector bias and flat-fielding effects, automatically measures the location of the guide star in each frame, and then shifts and aligns each frame to achieve an optimal image reconstruction using the Drizzle algorithm. During typical observing, where objects are generally scattered around the sky, we generally obtain residual wavefront errors in the 160 to 200 nm RMS range, leading to the ability to detect and characterize stellar companions at contrasts of >5 magnitudes at separations of 0.25-1" at visible wavelengths (see Figs. 3 and 4). The large 44" square field of view also allows for high-resolution imaging of extended objects such as Jupiter (Fig. 5).

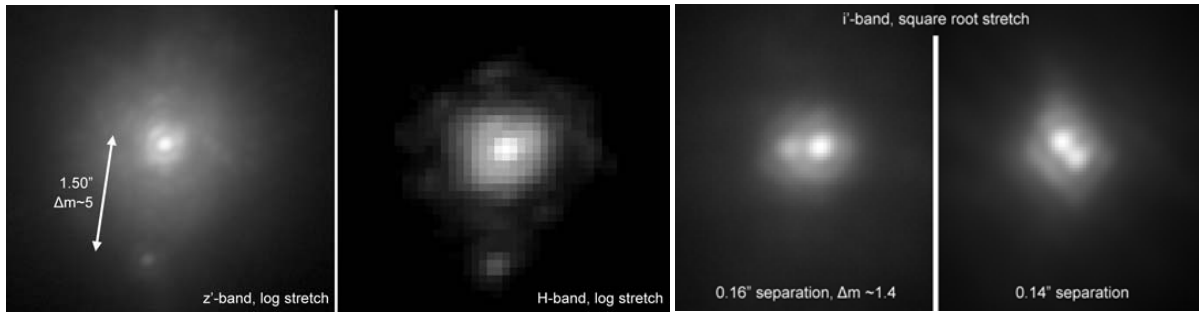
Fig. 3. Binary star systems imaged in the visible and near-infrared by Robo-AO.

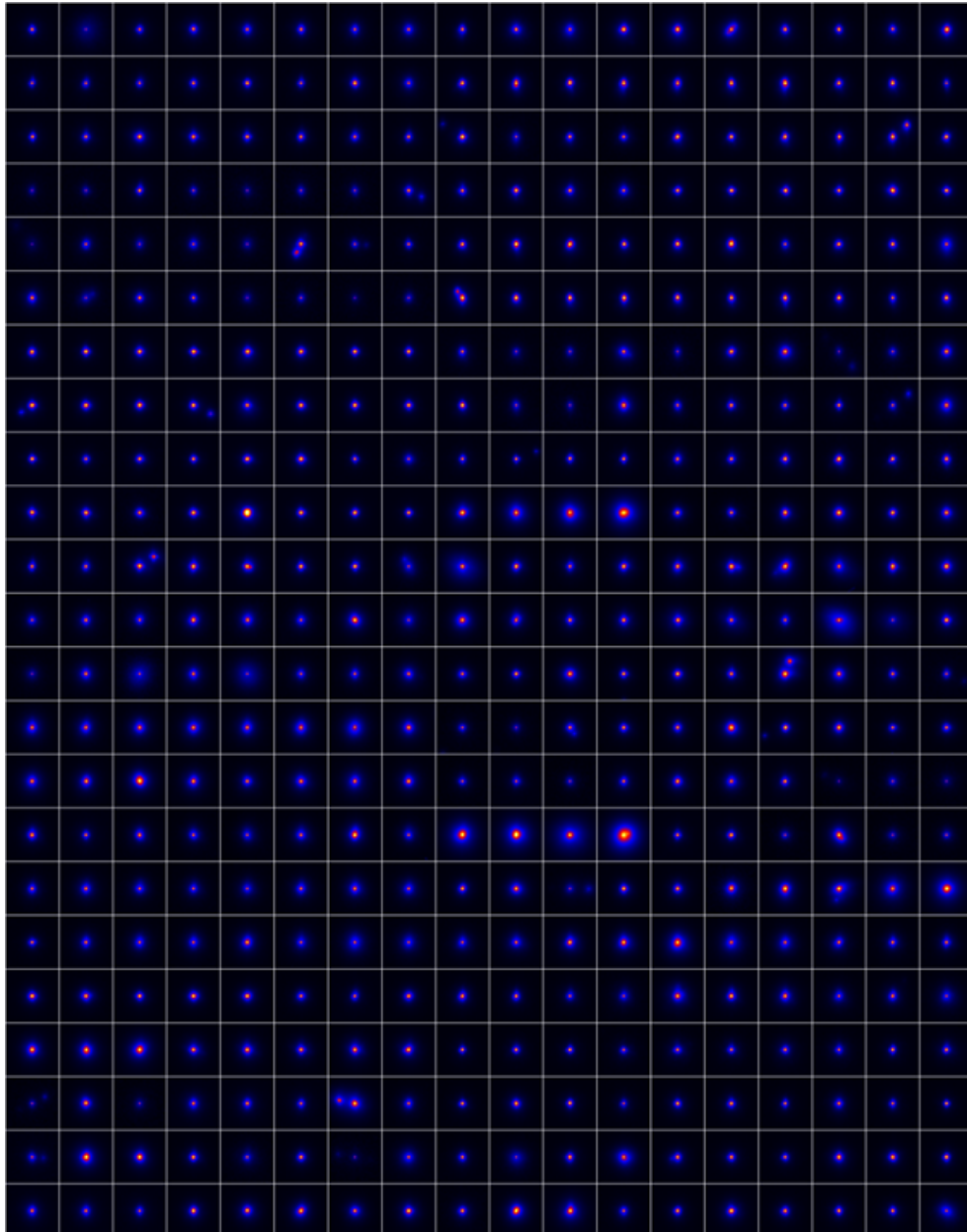
Fig. 4. Robo-AO adaptive optics images of 414 stars, executed in just 2.5 nights, as part of the 'Binarity of the solar neighborhood' project (Table 1). Each square represents a 3" × 3" area, and 90 s of integration in i-band ($\lambda = 765$ nm), sufficient to reach the photon noise floor in the halo and detect companions at 5-magnitudes within 0.2".

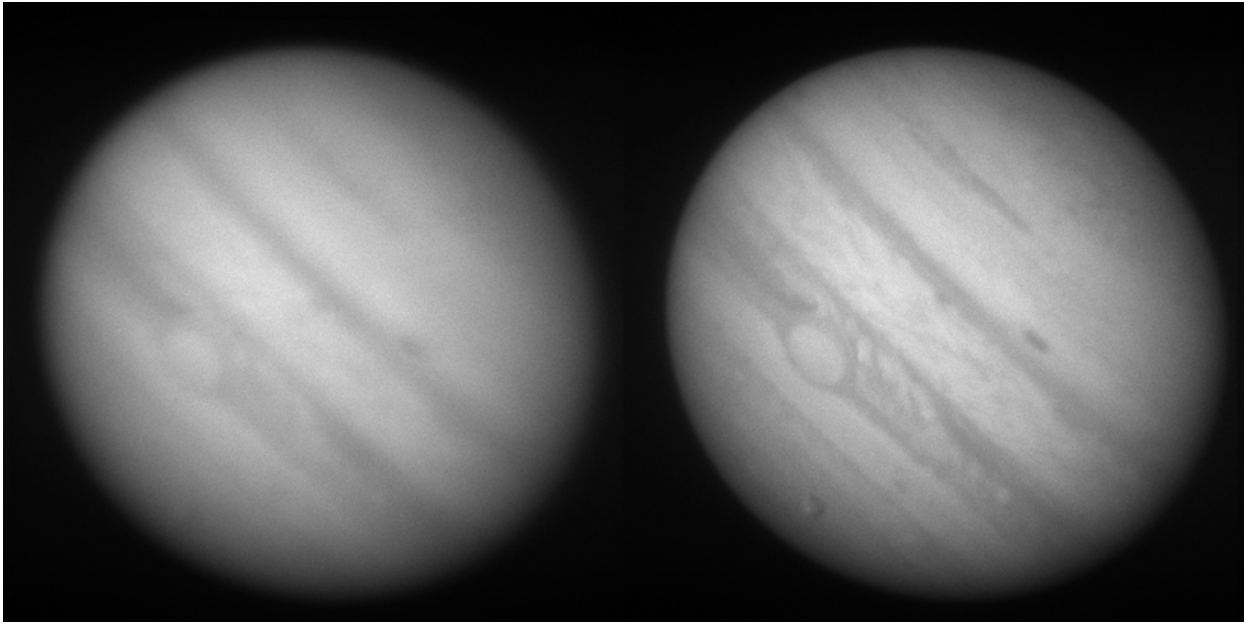

Fig. 5. Images of Jupiter. (Left) A 0.033-second uncompensated snapshot of Jupiter (apparent diameter of 42 arc seconds) in r-band (λ = 560 - 670 nm). (Right) The same image with Robo-AO laser adaptive optics correction showing the surface cloud features and transiting Ganymede with greater clarity (lower left of image).

## 4. ROBOTIC SOFTWARE AND AUTOMATIONS

The Robo-AO robotic control software [5] was developed in parallel with the hardware, to avoid hardware choices that would limit software functionality and to enhance the efficiency of the final system. A modular software architecture is employed that connects individual hardware components into software subsystems that run as daemons (e.g. the AO daemon operates the deformable mirror, wavefront sensor detector, image tip tilt motor, and laser uplink tip tilt motor). Each daemon is able to manage operation of its associated hardware automatically and react internally to hardware errors. Communications between the daemons use custom functions for command and control operations that pass continual status information and will automatically restart any daemons that lock up or crash. The daemons are controlled by a central robotic control daemon that commands observation sequences, monitors the state and health of the subsystems, reacts to errors found by the subsystems, and manages all aspects of operation of the observing system. The modular design isolates problems, so that even major issues like daemon crashes result in procedural steps to resume normal operations (instead of system crashes and lost hours of operation), which creates a robust, efficient automated system that can successfully complete scientific observations throughout the night.

As a fully autonomous laser AO instrument, Robo-AO executes tasks that are generally performed manually. The robotic system operates multiple subsystems in parallel in order to increase efficiency; for example, to start an observation, the central robotic control daemon will point the telescope, move the science filter wheel, and configure the science camera, laser, and AO system, all before the telescope has settled onto a new science target. The automated laser acquisition process begins once the telescope has been pointed. A spiral search algorithm acquires the laser by moving the uplink steering mirror until 80% of the wavefront sensor subapertures have met a flux threshold of 75% of the typical laser return flux. A safety system manages laser propagation onto the sky, and stops laser operations if any errors occur. Robo-AO uses a newly developed system for laser deconfliction that opens the entire area above 50 degrees zenith distance for observation by requesting predictive avoidance authorization for ~700 individual fixed azimuth and elevation boxes of ~6 square degrees every night. This gives Robo-AO the capability to undertake laser observations of any target overhead at almost any time, increasing observing efficiency by removing the need to preselect targets of observation. It should be noted that this method can be implemented for all laser AO observatories to reduce bookkeeping and make immediate target of opportunity observations possible.

While manually operated laser AO systems typically require 5-35 minutes to prepare for AO operation [11-13], Robo-AO requires ~50-60 seconds, on average, from the end of a telescope slew to the beginning of integration with the science camera; many nights of 200+ observations have already been achieved (with a current record of 240). As of this writing, Robo-AO has completed ~446 hours of fully robotic operations during 78 nights of allocated telescope time, of which ~242 hours (54%) were open-shutter science observing time. In total, ~9,000 observations were made, with typical exposure times ranging from 30 s to 3 minutes each, which comprise some of the largest high-angular resolution surveys ever performed, as indicated in Table 1. The Robo-AO "Binarity of the solar neighborhood" survey is a comprehensive study of the companions of nearby stars, < 35 pc, exploring the variation in their properties with stellar mass, metallicity, age and formation / multiplicity environment. The more than 3,000 observed targets are five times larger in number than any other non-Robo-AO adaptive optics survey to date. Robo-AO is able to complete these large surveys much faster than other AO systems, such as having observed ten times the number of Kepler Objects of Interest (KOI) than larger telescope facilities in half of the observing time. A third large survey of the multiplicity of solar type stars exceeding 1,000 targets has also been performed.

Table 1. Representative sample of the largest ground-based diffraction-limited surveys performed with telescopes greater than 1-m in diameter (not necessarily complete).

| Survey | Tel./Instrument | Method | Targets | Time | Reference |
|---|---|---|---|---|---|
| **Binarity of the solar neighborhood** | **P60 (Robo-AO)** | **LGS AO** | **3,081** | **172 hours** | **Law, et al. in prep.** |
| **Solar-type dwarf multiplicity** | **P60 (Robo-AO)** | **LGS AO** | **1,115** | **49 hours** | **Tokovinin, et al. in prep.** |
| M-Dwarf multiplicity | Calar Alto 2.2-m (AstraLux), NTT (AstraLux) | Lucky | 761 | 300 hours | Janson, et al. 2012 [14] |
| Washington Double Star Catalog | SOAR (HRCam) | Speckle, AO+Speckle | 639 | 16 nights | Hartkopf, et al. 2012 [15] |
| Young Solar analogs | KeckII (NIRC2), Hale (PHARO) | NGS AO | 266 | 47 nights | Metchev & Hillenbrand, 2009 [16] |
| Planets around low-mass stars | KeckII (NIRC2), Subaru (HiCIAO) | NGS AO | 125 | 12 nights | Bowler et al. 2012 [17], Bowler et al. in prep. |
| Gemini deep planet survey | GeminiN (NIRI) | NGS AO | 85 | 84 hours (w.o. overheads) | Lafrenière, et al. 2007 [18] |
| Multiplicity at the bottom of the IMF | KeckII (NIRC2) | LGS AO | 78 (87 obs.) | 10 nights | Kraus & Hillenbrand, 2012 [19] |
| **Kepler KOI validation** | **P60 (Robo-AO)** | **LGS AO** | **>1,700** | **104 hours** | **Law, et al. in prep.** |
| Kepler KOI validation | MMT (Aries), Hale (PHARO) | NGS AO | 90 | 12 nights | Adams, et al. 2012 [20] |
| Kepler KOI validation | Calar Alto 2.2-m (AstraLux) | Lucky | 98 | 19 nights | Lillo-Box, et al. 2012 [21] |

## 5. FUTURE WIDE-FIELD INFRARED CAMERA

The Robo-AO collaboration is currently in the process of upgrading the infrared imaging system from an engineering grade camera to a low-noise wide-field imager using a 2.5 μm cutoff Teledyne HAWAII-2RG™/HgCdTe detector (H2-RG) which was delivered to Caltech in September 2012. The science grade H2-RG detector has ~ 0.01 e-/sec dark current, < 5 e- readout noise [22] and slightly higher quantum efficiency than the InGaAs, offering 45-80 times more sensitivity (Table 2). The Robo-AO instrument has an external camera port (Fig. 6) designed to accommodate the expected 70-kg mass of the infrared camera. The optical design will consist of a simple reimaging system with cold stop and filters (J, H, K, H-br, CH4 on & off, a blank, and a spare) at the internal pupil plane. We have adopted a plate scale of 0.059"/pixel, Nyquist sampled at λ = 860 nm (roughly SDSS z-band), corresponding to a 2' diameter unvignetted field of view inscribing the square detector.

Furthermore, as part of our integration plan, we will develop automated routines for configuring the high-speed infrared tip-tilt sensing and recording of infrared science data in much the same way as has been done with the visible Robo-AO camera. Special procedures for dealing with infrared data, including taking sky-flats, backgrounds, and darks, as well as compensating for nonlinear responses and amplifier glow, will be developed and included as part of a standard data pipeline. We will also maintain a flexible FITS header for all data captured with Robo-AO's infrared and visible cameras, making it Virtual Observatory-compliant as recommended in the Astro-2010 Decadal survey for possible later public release and archiving.

Table 2. Optical format and infrared sensitivity of the Robo-AO InGaAs and H2-RG infrared cameras.

| Detector | Field of view | Plate Scale | J sensitivity[1] | H sensitivity[1] | K (λ=2.2 um) sensitivity[1] |
|---|---|---|---|---|---|
| InGaAs | 32.0"×25.6" | 0.10"/pixel | 14.9 | 15.2 | N/A |
| H2-RG | > 67"×67" | < 0.086"pixel | 19.7 | 19.4 | 18.2 |

[1] Faintest magnitude to reach a signal-to-noise ratio of 10 in 120 s of exposure. These assume a residual wavefront error of 185 nm RMS, expected AO performance under median seeing conditions.

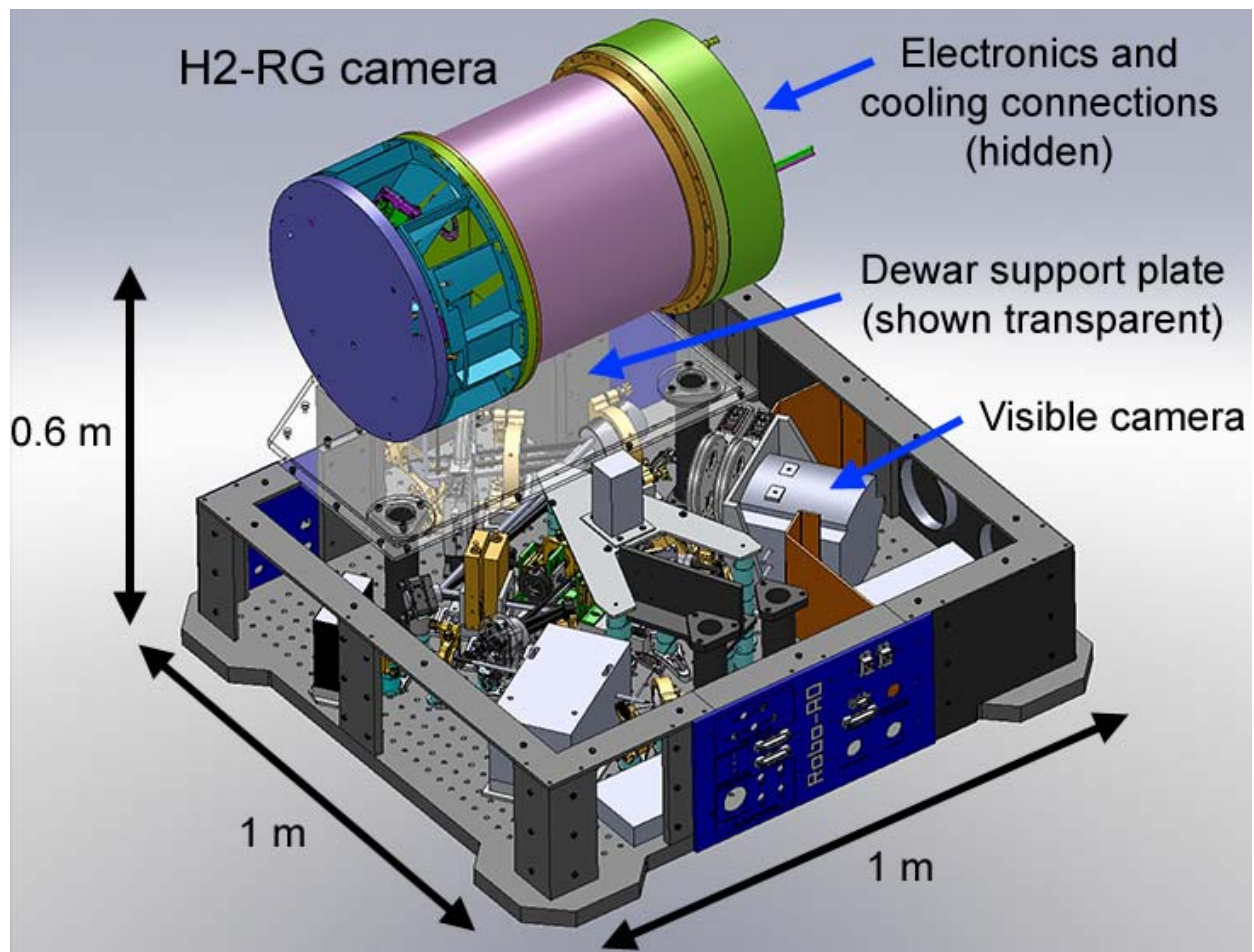

Fig. 6. Solidworks model of the as-built Robo-AO system with a model of the Keck infrared tip-tilt sensor (similar in function, size and weight to the planned Robo-AO infrared camera) attached.

## 6. CLONE OF ROBO-AO FOR IUCAA GIRAWALI OBSERVATORY

A clone of the current Robo-AO system is currently being developed for the 2-m IUCAA Girawali Observatory telescope [6] in Maharashtra, India. Adaptations in optical prescriptions have been made to accommodate the slightly different F/#, F/10 vs. F/8.75, different mounting interface and telescope interface software. No other major changes are anticipated - minimizing further development and control costs. The system is expected to see first light in 2014.

## 7. ALL-SKY ROBO-AO NETWORK

The prototype Robo-AO at Palomar has been crucial in validating the current sample of KOIs, having observed over three-quarters of the 2,036 host stars (from the January 2013 release), see Table 1. A complementary follow-on mission to Kepler is the Transiting Exoplanet Survey Satellite (TESS) led by MIT and scheduled for launch in 2017. TESS will execute a shallower survey compared to Kepler, with the majority of objects mV < 16, but over the entire sky, surveying the northern sky in year 1 and the southern sky in year 2. It is estimated that there may be as many as ten or more times as many transit signals discovered by TESS during its mission lifetime – which could all be validated by Robo-AO, in an analogous way to the KOIs, in less than a year. To make this happen, a North-South network of facility-class Robo-AO systems would be necessary: we are planning to deploy such systems at either or both of the 2.2-m UH and 3-m IRTF telescopes on Mauna Kea and are looking for partners for a facility-class Robo-AO in the southern hemisphere, preferably on a robotic 1.5 to 3+ m sized telescope.

## 8. ACKNOWLEDGEMENTS

The Robo-AO system is supported by collaborating partner institutions, the California Institute of Technology and the Inter-University Centre for Astronomy and Astrophysics, by the National Science Foundation under Grant Nos. AST-0906060 and AST-0960343, by a grant from the Mt. Cuba Astronomical Foundation and by a gift from Samuel Oschin. The infrared camera upgrade is additionally supported in part by the National Science Foundation under Grant No. AST-1207891 and by the Office of Naval Research under grant N00014-11-1-0903. We are grateful to the Palomar Observatory staff for their support of Robo-AO on the 60-inch telescope, particularly S. Kunsman, M. Doyle, J. Henning, R. Walters, G. Van Idsinga, B. Baker, K. Dunscombe and D. Roderick.

## 9. REFERENCES

1.  N. M. Law, et al., ApJ, 757, 133, (2012)
2.  P. Muirhead, et al., ApJ, 767, 111, (2013)
3.  J. Swift, et al., ApJ, 764, 105, (2013)
4.  E. Terziev, et al., ApJS, 206, 18, (2013)
5.  R. Riddle, et al., SPIE Adaptive Optics Systems III, 8447, 8447-96, (2012)
6.  R. Gupta, et al., Bulletin of the Astronomical Society of India, 30, 785, (2002)
7.  S. Severson, et al., SPIE MEMS Adaptive Optics VII, 8617, 8617-09, (2013)
8.  S. B. Cenko, et al., PASP, 118, 1396-1406, (2006)
9.  C. Baranec, et al., J. Vis. Exp., 72, e50021, doi:10.3791/50021 (2013)
10. N. M. Law, et al., ApJ, 692, 924-930, (2009)
11. Y. Minowa, et al., SPIE Adaptive Optics systems III, 8447, 8447-52, (2012)
12. B. Neichel, et al., SPIE Adaptive Optics systems III, 8447, 8447-176, (2012)
13. P. Wizinowich, et al., PASP, 118, 297-309, (2006)
14. M. Janson, et al., ApJ, 754, 44, (2012)
15. W. I. Hartkopf, A. Tokovinin & B. D. Mason, ApJ, 143, 42 (2012)
16. S. Metchev & L. A. Hillenbrand, ApJS, 181, 62-109, (2009)
17. B. P. Bowler, et al., ApJ, 753, 142, (2012)
18. D. Lafrenière, et al., ApJ, 670, 1367-1390, (2007)
19. A. L. Kraus & L. A. Hillenbrand, ApJ, 757, 141, (2012)
20. E. R. Adams, et al., ApJ, 144, 42, (2012)
21. J. Lillo-Box, D. Barrado & H. Bouy, A&A, 546, A10, (2012)
22. J. W. Beletic, et al., SPIE, 7021,7021-20, (2008)